\documentstyle[12pt]{article}
\newcommand{\be}{\begin{equation}}
\newcommand{\ee}{\end{equation}}
\newcommand{\ba}{\begin{eqnarray}}
\newcommand{\ea}{\end{eqnarray}}

\begin{document}
\renewcommand{\baselinestretch}{1.8}
\small\normalsize
\renewcommand{\theequation}{\arabic{section}.\arabic{equation}}
\renewcommand{\thesection}{\Roman{section}.}
\renewcommand{\thefootnote}{\fnsymbol{footnote}}
\language0

\title{Cluster mechanisms in the fully frustrated Ising model}
\author{Werner Kerler and Peter Rehberg}
\date{\sl Fachbereich Physik, Universit\"at Marburg, D-35032 Marburg,
Germany}
\maketitle
\begin{abstract}
The cluster algorithm in the fully frustrated Ising model on the square
lattice is essentially
different from the ones used in other systems. Thus its better understanding
is particularly important for finding new lines of development. Therefore we
investigate it in detail. In our simulations of high statistics more
appropriate choices of the probability for freezing bonds and of that for
flipping spins are seen to lead to better results. In an analysis of the
topological properties we derive a set of rules for possible cluster
configurations and give a classification by pairs of winding numbers.
\end{abstract}

\newpage
\renewcommand{\baselinestretch}{1.0}
\small\normalsize

\section{Introduction}
\hspace{0.35cm}
Cluster algorithms, introduced by Swendsen and Wang \cite{sw} for ferromagnetic
Potts models, allow a considerable reduction of critical slowing down in Monte
Carlo simulations. Beginning with $O(n)$-models \cite{wo} and $\phi^4$-theory
\cite{bt} various embeddings of the Ising dynamics have also been successful.
In the ferromagnetic Ising model in two dimensions forming blocks of spins
\cite{bt2} has been shown to lead to an improvement. The treatments of $Z_2$
gauge theories in three dimensions \cite{bh}, which use plaquette variables,
profit from the duality to the respective spin models. The combination of
cluster dynamics with making the strength parameter dynamical recently has led
to a very efficient method \cite{kw} to treat first order phase transitions in
the Potts model.

The algorithm for the fully frustrated Ising model on the square lattice in
two dimensions found by
Kandel, Ben-Av, and Domany \cite{kbd,kbd2} is of rather different type.
Therefore, insight in the respective mechanisms promises to show general rules
for the extension of cluster algorithms to other physical systems, which is
highly desirable, for example, for the gauge theories of particle physics or
for spin glasses. The understanding of this algorithm is, however, still
relatively vague. Therefore, it appears particulary important to investigate
it in more detail and to make the underlying features precise, which is the
purpose of the present paper.

We describe our Monte Carlo simulations of relatively high statistics. They
allow to determine the optimal choice of the freezing probabilities which is
important for analyzing the nature of the algorithm. The accuracy of the
numerical results makes it possible to discuss also details of the spectral
properties of the transition matrix. In particular a more appropriate flipping
rule for the cluster spins is found to lead to anticorrelations and thus to a
considerable improvement of the algorithm. Also interesting effects of the
lattice size dependence become obvious in this context. Furthermore, some
improvement of the associated Metropolis algorithm is possible.

We also analyze the topological properties of the clusters and and derive the
rules for the occurrence of specific cluster configurations. A
complete classification by pairs of winding numbers is given. The details of
the cluster-update mechanism become transparent. The better understanding
achieved in this analysis, apart from allowing to discuss the basic difference
to the other cluster algorithms in detail, indicates a new line of development.

In Sec.~II the simulations and their results are described. Sec.~III contains
some remarks on the assumptions entering the freezing probabilities and
Sec.~IV the spectral analysis. Sec.~V presents the topological analysis and the
discussion of its consequences.

\section{Simulations and numerical results}
\setcounter{equation}{0}
\hspace{0.35cm}
The fully frustrated Ising model is considered on a two-dimensional square
lattice with periodic boundary conditions. In its probability distribution of
spins $\mu(\sigma)\sim e^{-H(\sigma)}$ one has $H(\sigma)=
\sum_{\langle ij \rangle}(-J_{ij}\sigma_i\sigma_j+C_{ij})$ with the sum being
over nearest neighbor pairs, $\sigma_k$ taking the values $\pm 1$ , $C_{ij}$
specifying a normalization, and $|J_{ij}|=\beta$. To each plaquette one
antiferromagnetic and three ferromagnetic couplings are associated. Here the
antiferromagnetic links are chosen to be the vertical ones on even columns.

The model has a ferromagnetic critical
point at temperature $T=0$ (where $T=1/\beta$) \cite{vi,fg,wz}, which is of
primary interest here. To get a better overview also the region up to $T=1$ has
been investigated. Depending on lattice size and quantity considered the
results start to differ from the $T=0$ ones between $T=0.2$ and $0.6$.

To set up the cluster algorithm $H$ is decomposed into terms associated to
plaquettes, in particular to the shaded ones of a checkerboard pattern
\cite{kbd}. The freezing of bonds is related to the subdivision of the spin
variables on the plaquette into subsets (in which the variables are updated
simultaneously) as shown
in Fig.~1 and discussed below. On the plaquette either one or three links are
unsatisfied. The deleting probability, with which the freezing probabilities of
a plaquette sum to one, is taken to be exp$(H_{\alpha}(\sigma)-
\mbox{max}_{\{\sigma\}}(H_{\alpha}(\sigma))$, where $\alpha$ numbers the shaded
plaquettes and $H(\sigma)=\sum_{\alpha}H_{\alpha}(\sigma)$.

The cluster algorithm is not ergodic at $T=0$. In fact, we have observed
periods
between 2 and 200 sweeps. To get ergodicity also Metropolis updates have been
performed. It has turned out that using more than one Metropolis sweep per
cluster sweep is less favorable (the increase in CPU time needed is larger than
the small gain in autocorrelation time). Thus our full sweep consisted of one
sweep of each kind. This procedure has been maintained also at finite
temperature. To perform simulations directly at $T=0$ has been straightforward
(in Ref.~\cite{kbd2} $T=0.1$ was used which is close to this).

Our lattices in two dimensions with periodic boundary conditions have the
sizes $L=8$, 16, 24, 32, 40, 64, 128, 256. The observables measured are the
magnetization $M$ and the susceptibility $\chi$ (per site in each case). For
both of them also the autocorrelation functions have been determined. The
statistics collected is in the range of $1.2\times 10^6$ to $10^7$ sweeps.

In the choices of probabilities considered in Refs.~\cite{kbd} and \cite{kbd2}
and in our numerical simulations for freezing only one link can be unsatisfied.
The subsets frozen together in the choices P1, P2, and P3 used are depicted in
Fig.~1. From the numerical results in Refs.~\cite{kbd} and \cite{kbd2} and here
it is seen that P3 is to be excluded because (almost) the whole lattice is
frozen into one cluster. The prescription given in Ref.~\cite{kbd} can be
identified as a combination of P1 and P2 with a particular choice of weights
(for which a motivation has not been given). These weights have a
$T$-dependence
such that the case with P1 only is approached for $T\rightarrow 0$. In a later
paper \cite{kbd2} Kandel, Ben-Av, and Domany remarked that the probability
corresponding to P1 is a simpler legitime choice at finite temperature.

In our numerical work, checking systematically combinations of P1 and P2 with
respect to autocorrelations, we find that to use only P1 is optimal while to
use only P2 is worst. Combinations are found to be in between. The behavior of
the integrated autocorrelation times $\tau_{int}$ for P1 and P2 is illustrated
by Fig.~2. The curve between the extreme ones there is obtained for the
combination of probabilities introduced in Ref.~\cite{kbd} with the mentioned
particular weights.

The dependence of the autocorrelations on lattice sizes which
we find emphasizes our view that only P1 is appropriate. In the fit $kL^z$ to
$\tau_{int}$ we get for the dynamical critical exponent $z$ values between 0.22
and 0.35 (Table 1) for P1, however, about 1.8 for P2. Thus P2 shows the
behavior
of conventional algorithms and no reduction of critical slowing down.

It turns out that also in the Metropolis step, described by a proposal matrix
$W_p(\sigma;\sigma')$ and the acceptance matrix $\mbox{
min}\big(1,\mu(\sigma')/
\mu(\sigma)\big)$, some improvement is possible. At $T=0$ this step is not
ergodic if the Glauber choice $W_p=1$ is used and the spins within the
lattice are visited sequentially. To get ergodicity either $W_p=1$ with random
visits (as used \cite{ka} in Ref.~\cite{kbd2}) or $W_p=1-\frac{1}{2}\delta_
{\sigma,\sigma'}$ with sequential visits (as investigated here) may be chosen.
Both of these choices turn out to lead to about the same result for the
autocorrelation function. However, it is also possible to combine the
nonergodic
Metropolis step where $W_p=1$ with the nonergodic cluster step. This leads to
an ergodic algorithm with about 20\% smaller values of $\tau_{int}$.

There is considerable freedom for choosing
the flipping rules for the cluster spins which can be exploited to get optimal
ones \cite{k2,k3}. Thus, in addition to the one of Swendsen and Wang \cite{sw}
(SW) (where the new spins are independent of the old ones), we also have used
the one of flipping the largest cluster \cite{bc} (LC). For the flipping rule
LC
the integrated autocorrelation times turn out to be about a factor 3 smaller
than for SW. This remarkable result is caused by the occurrence of a negative
eigenvalue of the transition matrix, as will be discussed in more detail in
Sec.~IV. The related oscillating behavior of the autocorrelation function is
seen in Fig.~3.

It turns out that -- apart from very fast contributions only noticable close to
$t=0$~-- there is one mode of the transition matrix for the SW rule, however,
there are two modes for the LC rule. Thus $\chi^2$-fits to the autocorrelation
functions have been performed using the fit function $c\lambda^t$ for SW and
$c_1\lambda_1^t+c_2 \lambda_2^t$ for LC. The occurring eigenvalues are related
to the exponential autocorrelation times by $\lambda=e^{-1/\tau}$, $\lambda_1=
e^{-1/\tau_1}$ and $\lambda_2=-e^{-1/\tau_2}$, respectively. The integrated
autocorrelation times have been determined using the fit function for
extrapolating the autocorrelation function at larger $t$ \cite{k2}. Their error
has been determined making use of the vanishing of higher cumulants \cite{pr}.

To present quantitative results on the lattice size dependences, fits to
$kL^z$ have been performed for integrated and exponential autocorrelation times
and also for the coefficients describing the weights of the eigenvalues. Only
the sizes $L\ge 16$ have been included in these fits because of inacceptable
deviations for smaller lattices. The values obtained for $z$ and $k$ are listed
in Tables $1-3$. The errors given are statistical ones only. Implications of
these results will be discussed in Sec.~IV.

In our simulations also details of the cluster structure have been determined.
For P1 the number of clusters observed at $T=0$ is 2 in about 96\% of cases,
4 in most of the rest, and very rarely 6. The mean relative size of the largest
cluster is 0.6435(5) independently of lattice size. On the other hand, for P2
there are about 19 clusters (one large and the other ones small) and the mean
size of the largest cluster is about 0.92.

Tables 4 and 5 present our detailed results on the cluster properties for P1 at
$T=0$. In addition to data on cluster numbers $N_c$ also ones on
characteristic pairs of winding numbers $\bf w$ are given. These quantities
will be explained and discussed in Sec.~V.

The high statistics collected and the reduction of critical slowing down
achieved also allow to get precise results (much more accurate than those of
Ref.~\cite{kbd2}) on the scaling properties of the observables $M$ and $\chi$
at
$T=0$. The fits $k_M L^{z_{M}}$ and $k_{\chi}L^{z_{\chi}}$, respectively, give
the results for $z_M$ and $z_{\chi}$ shown in Fig.~4 for the variants of the
algorithm considered \cite{stat}. They are plotted against the size of the
smallest lattice to which the particular fit extends in order to show the
impact of finite size effects. The errors given are statistical ones (and thus
are not available at the highest-$L$ points). It is seen that within errors
(statistical ones plus expected systematic ones) the theoretical values
$z_M=-0.25$ and $z_{\chi}=1.5$
are approached. The corresponding $k$ values are $k_M$=0.678(1) and $k_{\chi}$=
0.580(2). The relation $z_{\chi}-2z_M=2$ is satisfied for $L\ge 16$ within
statistical errors.

\section{Assumptions about transition probabilities}\setcounter{equation}{0}
\hspace{0.35cm}
A systematic derivation of the choices P1, P2, and P3 of the freezing
probabilities reveals that there is quite a number of assumptions which
enter. We briefly summarize our results within this respect here which
appear useful for further work.

To get stationarity it is convenient (though not necessary) to impose detailed
balance. To see how far this restricts the choices of the probabilities one
needs an appropriate formulation. In the general framework given by Kandel and
Domany \cite{kd} the construction of a new Hamiltonian is equivalent to
introducing a joint probability distribution $\mu_J(n,\sigma) =
\mu(\sigma)A(\sigma;n)$ where $\mu$ is the given distribution and
where $A$ is the conditioned probability for getting a configuration of the
newly introduced variables $\{n\}$ if a configuration $\{\sigma\}$ is given.
Their condition for detailed balance then amounts to require
$\mu_J(n,\sigma)B(n,\sigma;\sigma') =\mu_J(n,\sigma')B(n,\sigma';\sigma)$
where $B(n,\sigma;\sigma')$ is the conditioned probability to get $\{\sigma'\}$
if both $\{\sigma\}$ and $\{n\}$ are given. Inserting the definition of $\mu_J$
and noting that the transition probability of interest is of form
$W(\sigma;\sigma')=\sum_{\{n\}} A(\sigma;n) B(n,\sigma;\sigma')$ it becomes
obvious that detailed balance already arises by summing their condition.

Thus to get restrictions on the probabilities more detailed specifications
\cite{k0} are necessary. In the form $A(\sigma;n)
=\prod_{\alpha}p_{\alpha}(\sigma;n_{\alpha})$, where $\alpha$ is related to
the decomposition $H(\sigma)=\sum_{\alpha}H_{\alpha}(\sigma)$, in the present
application the $p_{\alpha}$ describe freezing for $n_{\alpha}>0$ and
deleting for $n_{\alpha}=0$. Then one has to put
$p_{\alpha}(\sigma;0)=\mbox{exp}(H_{\alpha}(\sigma)-
\mbox{max}_{\{\sigma\}}(H_{\alpha}(\sigma))-\Delta_{\alpha})$,
where $\Delta_{\alpha}\ge 0$ are arbitrary constants, and to impose certain
conditions \cite{k0} on $B$ in order to get detailed balance. The freedom
in the choice of $B$ allows to optimize the algorithm by finding more
appropriate flipping rules \cite{k2,k3}. For the present purpose these
conditions \cite{k0} provide the desired restrictions of probabilities.

In this context it is to be noted that for the particular flipping rule of
Swendsen and Wang \cite{sw} (SW), the transition matrix simplifies to
$\tilde{W}(\sigma;\sigma')=\sum_{\{n\}}A(\sigma;n)\tilde{A} (n;\sigma')$
where $\tilde{A}(n;\sigma)= \mu_J(n,\sigma)/\sum_ {\{\sigma\}}\mu_J(n,\sigma)$.
The consequence is \cite{k0} that for the SW rule detailed balance follows
immediately without further assumptions. Thus, because in Refs.~\cite{kbd,kbd2}
only the SW rule is used, to show detailed balance the proof as given in
Ref.~\cite{kbd2} is actually not necessary. To restrict the probabilities also
in the SW case the restrictions obtained for general flipping rules are to be
imposed by assumption.

The plaquette terms into which $H$ of the fully frustrated model is decomposed
are of form $-4\beta\sum_{\alpha}(\delta_{\alpha}(\sigma)+c_{\alpha})$ where
$\delta_{\alpha}(\sigma)=1$ if three links are satisfied and $\delta_{
\alpha}(\sigma)=0$ if only one link is satisfied (with irrelevant constants
$c_{\alpha}$). Thus as in the ferromagnetic Potts case \cite{sw} only two
values
are taken, parallel neighboring spins there corresponding to three satisfied
links in a plaquette here.

To get restrictions for the occurring probabilities we use the conditions
\cite{k0} which apply for general flipping rules. In $p_{\alpha}(\sigma;0)$
the additional requirement $\Delta_{\alpha}=0$ is imposed (which we have
checked
numerically to be optimal in the fully frustrated as well as in the ordinary
Ising model). A further additional condition is that the
$p_{\alpha}(\sigma;n_{\alpha})$ for $n_{\alpha}>0$ have the form
$\delta_{\alpha}(\sigma)\hat{p}_{\alpha n_{\alpha}}$ with $\hat{p}_{\alpha
n_{\alpha}}\ge 0$ being independent of $\{\sigma\}$. Still further
assumptions are needed to fix the freezing probabilities, the possibilities for
which are related to partitions of the spin variables on the plaquette into
sets.

To motivate additional restrictions we observe that the condition
$\delta_{\alpha}(\sigma')=\delta_{\alpha}(\sigma)$ for $n_{\alpha}>0$,
which with the above assumptions follows for $\delta_{\alpha}(\sigma)=1$
{}from the general requirements \cite{k0},
is not respected in all of the partitions of variables for all
possible updates of the spins on the plaquette (in which cases the general
properties \cite{k0} of $B$ prevent them from contributing). This suggests
to simplify things by excluding all partitions where this can happen. If this
prescription is adopted one remains with the three possibilities P1, P2, and P3
considered in the numerical work. Thus quite a number of assumptions and much
stronger restrictions than necessary are required to arrive at these freezing
probabilities.

\section{Effects of spectral properties}
\setcounter{equation}{0}
\hspace{0.35cm}
To discuss the spectral properties of the transition matrix $W(\sigma;\sigma')$
it is convenient to introduce an inner product \cite{msls} with respect to
which
detailed balance for $W$ is just the condition for self-adjointness. The
generel relations applying in the present context have been given recently
\cite{k3}.

In the application here detailed balance is satisfied by the cluster steps
and by the local Metropolis steps but not by the product of the respective
matrices. One has, however, to note that for $W$ in $(f,Wf)$ instead of a
product $W_1 W_2 \ldots W_M$ one can equivalently use $\frac{1}{2}(W_1 W_2
\ldots W_M + W_M \ldots W_2 W_1)$ which for $W_i^{\dagger}=W_i$ is
self-adjoint.
Thus with the appropriate identification one gets self-adjointnes also for $W$.

The spectral representation of the normalized autocorrelation function
$\rho(t)=R(t)/R(0)$, where $R(t)=
\langle f_s f_{s+t}\rangle-\langle f_s\rangle^2$, has the form
\be
\rho(t)=\sum_{\nu}\lambda_{\nu}^{\,t}\:c_{\nu}
\label{4e31}
\ee
(in which the eigenvalue 1 of $W$ does not occur).
{}From (\ref{4e31}) and $\rho(0)=1$ one gets the condition
$\sum_{\nu}c_{\nu}=1$ for the coefficients. Thus, if detailed balance holds, in
which case all $c_{\nu}$ must be positive, they also cannot exceed 1~. Further,
then the eigenvalues are real and the parametrizations $\lambda_{\nu} = e^{-1/
\tau_{\nu} }$ and $=-e^{-1/\tau_{\nu} }$ are appropriate.
The integrated autocorelation time is related to $\rho$ by
\be
2\tau_{int}= 1+2\sum_{t\ge 1}\rho(t)   \quad .
\label{4e31a}
\ee
The factor $\sqrt{2\tau_{int}}$, which  enters the statistical error, should
be made as small as possible.

For the flipping rules SW and LC, considering the observables $M$ and $\chi$
which depend on the spins only, one has detailed balance (for an example where
only stationarity holds see Ref.~\cite{k3}). For SW in addition only positive
eigenvalues occur \cite{k3}.  This is, however, not desirable as inspection of
(\ref{4e31a}) with (\ref{4e31}) shows. Apart from the eigenvalue 1 one would
rather like to have a negative spectum which leads to a reduction of the error
by anticorrelations. For LC positive as well as negative eigenvalues are
allowed
and, as already reported in Sec.~II, indeed occur.

{}From Table 3 it is seen that (similarly as in the ferromagnetic Ising case
\cite{k3}) also the coefficients in (\ref{4e31}) depend on the lattice size.
It is seen that this dependence contributes significantly to that of the
integrated autocorrelation times. In principle the fact pointed out above that
the $c_{\nu}$ cannot exceed 1 causes deviations from a power law. However, in
the $L$ range considered in our fit the values are well below 1 and the
deviations would occur only at unrealistic small lattices.

It should be noted that at finite values of $L$ simultaneous power-law
behaviors
of $\tau_{int}$, $\tau_{\nu}$, and $c_{\nu}$ cannot be expected in general.
This
occurs because a term $c_{\nu}\lambda_{\nu}^{\,t}$ in (\ref{4e31}) gives the
contribution $2c_{\nu} (\lambda_{\nu}^{-1} -1)^{-1}$ to (\ref{4e31a}). Thus
only
if a term with a positive $\lambda_{\nu}$ dominates in (\ref{4e31a}) and
$\tau_{\nu}$ is sufficiently large this contribution approaches $2c_{\nu}\tau_
{\nu}$, such that the laws $k_{c\nu}L^{z_{c\nu}}$ and $k_{\tau\nu}L^{z_ {\tau
\nu}}$ can combine to the simultaneous law $k_{c\nu} k_{\tau\nu}
L^{z_{c\nu}+z_{\tau\nu}}$.
In the cases studied here the conditions for simultaneous power laws are not
met mainly because the values of $c_1$ are rather small and the required
dominance is not there. Nevertheless within errors power-law fits are
convenient to represent the data. Further, empirically one observes that
${z_{c\nu}+z_{\tau\nu}}$ is rather close to $z$ of $\tau_{int}$ for $\nu=1$ and
also for $\nu=2$.

For LC $c_2$ is much larger than $c_1$ such that the negative mode is most
important. The fact that the relative contribution of the positive mode
increases with lattice size does not matter at the lattice sizes which can be
reached in practice. The effective weight $1-c_1-c_2$ of very fast
contributions
(only visible close to $t=0$) is seen to be not small. These contributions must
contain positive spectral parts because otherwise the values of $\tau_{int}$
would be much smaller.

An important observation here is that there are different lattice-size
dependences for the eigenvalues as well as for the weights of different
modes. With respect to devicing algorithms the fact that different dependences
are possible means that there are no general restrictions which, for example,
would forbid the optimal case of having only negative eigenvalues in addition
to the value 1.

\section{Cluster mechanism and winding numbers}
\setcounter{equation}{0}
\hspace{0.35cm}
The analogy of parallel neighboring spins in the ferromagnetic Ising case to
three satisfied links on a plaquette here has been mentioned in Sec.~III . The
only difference at the formal level is that in the fully frustrated case there
are more freezing possibilities. Considering $T=0$ there is, however, also an
essential difference from the dynamical point of view. There is no analogue of
domains in which one could have finite growths of clusters. Thus it is obvious
that the usual cluster mechanism (of filling part of a domain and thus breaking
extended structures) cannot work. Therefore, it could have been foreseen that
using the partition P3 must fail.

The working of their algorithm has been explained by Kandel, Ben-Av, and Domany
by the occurrence of at least two clusters of length $L$ \cite{kbd}. Giving a
proof in terms of the dual lattice and of sublattices thereof in
Ref.~\cite{kbd2} they conclude that at least two nontrivial loops and no
trivial
loop are created. In the following we derive detailed properties of the cluster
configurations which occur at T=0 using the algorithm with the partition P1.

The key observation allowing an easy analysis is that there is a simple
description of the ground states between which the transitions in equilibrium
take place. The basic building blocks of these states are pairs of plaquettes
with three satisfied links having the unsatisfied link in common, as is shown
in
Fig.~5a. The ground states then are obtained by tiling the lattice with these
building blocks, as is illustrated by Fig.~5b.
In this context it should be mentioned that local updates are only able to flip
a spin surrounded by four plaquettes which belong to two basic blocks, which
gives rise to the transition depicted in Fig.~6.

After introducing shaded plaquettes of a checkerboard pattern, corresponding to
the decomposition of $H$, for the partition P1 one gets the rules for putting
clusters on the boundaries of the basic building blocks shown in Fig.~7.
Clusters are mandatory on the shaded-side parallel links and forbidden on the
shaded-side perpendicular link, while on the unshaded-side links clusters are
allowed and occur according to the situation in the neighboring building
blocks.

The clusters form closed nonintersecting paths. This follows immediately
considering the three ways a second building block can be attached to the
unshaded link next to a cluster-occupied shaded link of a first building block,
which are shown in Fig.~8. From this the only ways to continue the mandatory
cluster of the first block are obvious.

There are only topologically nontrivial clusters, i.e.~ones closing accross the
periodic boundaries. Trivial ones which may be contracted to a point by smooth
deformation are not possible. To see this one has first to note that the torus
may be described by an atlas of charts (mappings of a set of the manifold to a
set of ${\bf R}^2$) and that for a trivial loop a chart exists which contains
it
in its interior. Thus it remains to show that a loop in ${\bf R}^2$ cannot
exist.

Suppose there would be a loop in ${\bf R}^2$. Then it either must not enclose
further mandatory links or it must enclose a smaller loop (remember that the
loops are nonintersecting and behave on the basic blocks as indicated in
Fig.~7). Thus finally there must be a smallest closed loop no longer enclosing
mandatory links. The smallest closed loop then must enclose a set of basic
building blocks in such a way that the forbidden side of one block is attached
to an allowed one of another block (as illustrated by simple examples in
Fig.~9). For a set of $k$ blocks, however, this is only possible $k-1$ times.
Therefore, there always remains one forbidden link which prevents the loop from
closing.  Thus a contradiction is reached.

To characterize a nontrivial cluster on the two-dimensional torus one needs a
pair of winding numbers, ${\bf w}=(w_1,w_2)$. Empirically the combinations
${\bf w}=(1,w)$ and ${\bf w}=(w,1)$ with $w=0,1,2,3,\ldots$ illustrated in
Fig.~10 are observed (Table 5). To see more general possibilities we consider
the case of $n$ and $m$ nontrivial crossings in direction 1 and 2,
respectively.
For positive $n$ and $m$ any set of nonintersecting nontrivial loops can be
smoothly deformed to the standard form for which (up to a reflection of one
of the axes) Fig.~11 shows typical examples. The topology exhibited in the
standard form reveals that the largest common divisor $q$ of $n$ and $m$ gives
the number of clusters and that all of them have the winding numbers
${\bf w}=(n/q,m/q)$.

To establish these features for general positive integers $n$ and $m$ we use a
numbering of the nontrivial crossings in one direction. Without restricting
generality using direction 1 with the numbering $c_2=0,1,\ldots,n-1$ it follows
that starting
at a number $\tilde{c}$ and running along the loop after $s$ steps one arrives
at the number $\tilde{c}+sm$ mod $n$. To get back to $\tilde{c}$ therefore
requires
\be
sm=\lambda n
\label{sm}
\ee
where $\lambda$ is a positive integer. From (\ref{sm}) it is seen that if $q$
is the largest common divisor of $n$ and $m$ one gets back if $s=n/q$ and
$\lambda=m/q$. Then the number of clusters is $q$ and the winding numbers are
$w_1=n/q$ and $w_2=m/q$ for all of them (because the above arguments hold for
any choice of $\tilde{c}$).

If $n$ is zero the standard-form topology consists of $m$ lines in direction 2.
This means that one has $m$ clusters with winding numbers ${\bf w}=(0,1)$.
Analogously for $m=0$ there are $n$ clusters with ${\bf w}=(1,0)$.

Collecting the respective results we now see that a cluster can have the
winding numbers ${\bf w}=(w_1,w_2)$ with positive $w_1$ and $w_2$ which have
no common divisor larger than one. Further there is the possibility that one
of the winding numbers is zero in which case the other one must be one. It also
turns out that within any configuration all clusters must have the same pair
of winding numbers.

To find out which numbers of clusters $N_c$ are posssible in a configuration
it is to be noted that according to the above rules at least one winding number
must be odd. Because the extension of the lattice $L$ is even, an even number
of
clusters is needed to fill the boundary in the direction of the odd winding
number, i.e.~only the cluster numbers
\be
 N_c= 2, 4, 6, \ldots
\label{6e2}
\ee
are allowed, provided the cluster paths running towards the boundary and back
add only even numbers, which is shown in the following.

To study the paths running forth and back we first note that a path with pieces
of one lattice space apart must somewhere be connected as shown in Fig.~12a,
while one with pieces two spaces apart must have somewhere a connection as
shown in Fig.~12b in order to avoid the forbidden links of the procedure. This
readily generalizes to pieces an odd number of spaces apart and pieces an even
number of spaces apart, respectively. A path ultimately changing direction
must contain an odd number of parts of the odd-spacing type and can contain
any number of parts of the even-spacing type. Thus it runs back
after an odd number of spacings and therefore covers an even number of sites
at the boundary considered, independently of the particular behavior there.

The general laws on numbers of clusters and pairs of winding numbers derived
here are nicely confirmed by our numerical results. Tables 4 and 5,
respectively, show that larger numbers of clusters $N_c$ and larger winding
numbers $w$ are observed less frequently. For larger lattices this effect gets
stronger for $N_c$ and weaker for $w$. It turns out that only the pairs of
winding numbers of type ${\bf w}=(w,1)$ and ${\bf w}=(1,w)$ are observed in
practice while the occurences of ${\bf w}=(w_1,w_2)$ with $w_1$ and $w_2$ both
being larger than one are smaller than $10^{-6}$.

The description and the rules given here also allow to get the details of the
cluster-update mechanism. It turns out that (for fixed shaded plaquettes) there
are transitions between two ground states which occur by a kind of
``pipelining'' of basic blocks between two clusters, the spins in one of which
are flipped. This is illustrated in Fig.~13 for different types of cluster
configurations. In addition, it is seen in Fig.~13c that if a cluster forms
``spikes'', the blocks contained in a ``spike'' are not moved. It should
be obvious that the features found in the simple examples of Fig.~13 are the
general ones which occur on any lattice filled by nonintersecting clusters
(Fig.~9 provides more general examples of spikes).

In this context the effect of a local update (illustrated in Fig.~6) on a
subsequent cluster update is important. Fig.~14 gives examples that depending
on the result of the preceding local update one can get cluster configurations
with a different number of clusters $N_c$ (Fig.~14a) as well as with different
pairs of winding numbers ${\bf w}$ (Fig.~14b).

The above analysis shows that in the type of cluster algorithm considered
topological constraints are entirely responsible for the working of the
approach. This concept differs fundamentally from that of the finite growths of
clusters in domains used in the conventional approaches. Thus the suggestion
emerges to try this new line of development also in other cases where
conventional cluster methods fail.

In further developments of this direction one has to be aware that the
constraints found depend sensitively on the boundary conditions and on the
lattice type \cite{rich}. For example, using twisted boundary conditions the
mechanism does no longer work. Similarly, e.g. on a triangular lattice no
straightforward analogue is in sight. Thus some new strategies remain to be
worked out to utilize this concept on a broader basis. Nevertheless it is
attractive enough to warrant the respective effort.

\section*{Acknowledgements}
\hspace{0.35cm}
One of us (W.K.) wishes to thank Claudio Rebbi and the Physics Department
of Boston University for their kind hospitality during a sabbatical leave.
He also thanks Richard Brower for discussions on possible further developments.
This work has been supported in part by the Deutsche Forschungsgemeinschaft
through grants Ke 250/7-1 and Ke 250/9-1. The computations have been done on
the SNI 400/40 of the Universities of Hessen at Darmstadt and on the Convex
C230 of Marburg University.

\section*{Note added}
\hspace{0.35cm}
After completion of this work we obtained the Syracuse preprint SCCS-527
by P.D.~Coddington and L.~Han which also addresses issues of the cluster
algorithm in the fully frustrated Ising model.

\newpage
\renewcommand{\baselinestretch}{1.2}

\newpage
\renewcommand{\baselinestretch}{1.0}
\small\normalsize

\begin{center}

{\bf Table 1}

Fit to $kL^z$ for integrated autocorrelation times at $T=0$.

\begin{tabular}{|c|c|c|c|c|c|}\hline
&      & \multicolumn{2}{c|}{$M$}  & \multicolumn{2}{c|}{$\chi$} \\ \cline{3-6}
flip & $W_p(\sigma,\sigma)$
            &    $z$     &    $k$     &    $z$     &    $k$    \\ \hline
SW   & 0.5  &  0.236(3)  &  0.521(4)  &  0.281(2)  &  0.475(3) \\
SW   & 1    &  0.225(2)  &  0.440(3)  &  0.259(2)  &  0.405(3) \\
LC   & 1    &  0.311(9)  &  0.109(4)  &  0.355(8)  &  0.102(4) \\ \hline
\end{tabular}

\vspace{2.5cm}

{\bf Table 2}

Fit to $kL^z$ for exponential autocorrelation times at $T=0$.

\begin{tabular}{|c|c|c|c|c|c|c|}\hline
&  &  & \multicolumn{2}{c|}{$M$}  & \multicolumn{2}{c|}{$\chi$} \\ \cline{4-7}
flip & $W_p(\sigma,\sigma)$ & mode
            &    $z$     &    $k$     &    $z$     &    $k$    \\ \hline
SW   & 0.5  & 1 &  0.669(30) &  0.312(35)  &  0.716(22)  &  0.265(22) \\
SW   & 1    & 1 &  0.707(18) &  0.131(9)   &  0.712(16)  &  0.130(8)  \\
LC   & 1    & 1 &  0.88(8) &  0.053(15)  &  0.88(6)  &  0.055(13) \\
     &      & 2 &  0.459(8) &  0.430(11)  &  0.466(9)  &  0.417(12)  \\ \hline
\end{tabular}

\vspace{2.5cm}

{\bf Table 3}

Fit to $kL^z$ coefficients of modes at $T=0$.

\begin{tabular}{|c|c|c|c|c|c|c|}\hline
&  &  & \multicolumn{2}{c|}{$M$}  & \multicolumn{2}{c|}{$\chi$} \\ \cline{4-7}
flip & $W_p(\sigma,\sigma)$ & mode
            &    $z$     &    $k$     &    $z$     &    $k$    \\ \hline
SW   & 0.5  & 1 & -0.34(5)  &  0.58(11)  & -0.34(4) &  0.68(9)  \\
SW   & 1    & 1 & -0.501(24)  &  2.02(20) & -0.451(20) &  1.84(16) \\
LC   & 1    & 1 & -0.56(12) &  1.00(48) & -0.46(11) &  0.78(32) \\
     &      & 2 & -0.183(12)  &  1.31(6)  & -0.191(13) &  1.28(6)  \\ \hline
\end{tabular}

\newpage
\renewcommand{\baselinestretch}{1.0}
\small\normalsize

{\bf Table 4}

Occurrence of cluster numbers $N_c$ in \%

\begin{tabular}{|c|c|c|c|c|}\hline
        &  $N_c=2$  &  $N_c=4$  &  $N_c=6$  \\ \hline
$L=16$  &  96.5375(2)   &  3.4426(2)    &  0.0099(1)   \\
$L=32$  &  96.8846(3)   &  3.1083(2)    &  0.0071(1)   \\
$L=64$  &  97.0038(12)  &  2.9924(12)   &  0.0038(1)   \\ \hline
\end{tabular}

\vspace{2.5cm}

{\bf Table 5}

Occurrence of winding numbers $w$ in \%\\
within pairs ${\bf w}=(w,1)$ and ${\bf w}=(1,w)$\\

\begin{tabular}{|c|c|c|c|c|}\hline
        &  $w=0$  &  $w=1$  &  $w=2$  &  $w=3$  \\ \hline
$L=16$  &  83.34(1)    &  16.38(1)    &  0.276(2)    &  0.00012(5)  \\
$L=32$  &  82.77(5)    &  16.93(5)    &  0.304(4)    &  0.00021(8)  \\ \hline
\end{tabular}

\end{center}

\newpage
\renewcommand{\baselinestretch}{1.2}
\small\normalsize

{\Large \bf Figure captions}

\vspace*{0.5cm}

\begin{tabular}{rl}
Fig. 1. & Partitions of basic plaquette in cluster update\\
        & (unsatisfied link denoted by u).\\
Fig. 2. & $\tau_{int}$ for $\chi$ as function of $T$
          (errrors smaller than symbols)\\
        & for cluster updates with partitions P1 and P2,\\
        & and using prescription of Ref.~\cite{kbd} marked by KBD.\\
Fig. 3. & Typical autocorrelation function for the flipping rule LC \\
        & (shown for $\chi$ and $L=64$, with errors smaller than symbols).\\
Fig. 4. & $z_M$ and $z_\chi$ with statistical errors for rules SW with\\
  & $W_p(\sigma,\sigma)$=0.5 (dots), SW with $W_p(\sigma,\sigma)$=1
(crosses),\\
        & and LC (circles) as functions of smallest lattice in fit.\\
Fig. 5. & (a) Basic building block of ground states for $T=0$ (unsatisfied\\
        & link denoted by u). (b) Tiling of the lattice by basic blocks.\\
Fig. 6. & Effect of a local update on basic blocks (spin up denoted by point,\\
        & spin down by circle).\\
Fig. 7. & Rules for cluster location on basic block with shaded plaquette \\
        & specified (cluster mandatory m, forbidden f, allowed a).\\
Fig. 8. & Possibilities to attach upper block to lower block in order to\\
        & continue mandatory cluster of lower block (fat lines are clusters).\\
Fig. 9. & Examples showing that trivial closing is impossible.\\
        & (fat lines are clusters, f denotes forbidden links).\\
Fig.10. & Examples of pairs of winding numbers for one cluster\\
        & on lattice with periodic boundary conditions.\\
Fig.11. & Examples of the standard form of cluster topologies,\\
        & (a) with $n=5$ and $m=4$ (one cluster with ${\bf w}=(5,4)$)\\
        & (b) with $n=6$ and $m=4$ (two clusters with ${\bf w}=(3,2)$)\\
Fig.12. & Basic connections of pieces of loops (fat lines),\\
        & (a) if one spacing apart, (b) if two spacings apart.\\
Fig.13. & Cluster updates on periodic $4\times 4$ lattice for various\\
        & cluster configurations; (a) with ${\bf w}=(0,1)$ and $N_c=4$;\\
 & (b) with ${\bf w}=(1,1)$ and $N_c=2$; (c) with ${\bf w}=(0,1)$ and
$N_c=2$.\\
Fig.14. & Examples of cluster configurations occurring after a local update;\\
        & (a) $N_c=4$ or $N_c=2$ with ${\bf w}=(1,0)$; \\
        & (b) ${\bf w}=(1,0)$ or ${\bf w}=(1,1)$ with $N_c=2$.\\
\end{tabular}


\begin{thebibliography}{99}
\bibitem{stat} The static exponents $z_M$ and $z_{\chi}$ here should not be
mixed up with dynamical critical exponents related to autocorrelations.
\bibitem{rich} This has become clear in discussions with Richard Brower.
\end{thebibliography}
\end{document}